\documentclass[epj,twocolumn]{webofc}
\usepackage[varg]{txfonts}   
\usepackage{epsfig}
\usepackage{graphics}
\usepackage{graphicx}
\usepackage{multirow}
\woctitle{CGS15}
\begin{document}
\title{Microscopic nature of the photon strength function: stable and unstable Ni and Sn isotopes}
%
%

\author{Oleg Achakovskiy\inst{1}
\and
        Alexander Avdeenkov\inst{1}
              \and
        Stephane Goriely\inst{2}
             \and
             Sergei Kamerdzhiev\inst{3}\fnsep\thanks{\email{kaev@obninsk.com}} 
             \and
             Siegfried Krewald\inst{4}
             \and
             Dmitriy Voitenkov\inst{1}}

\institute{Instutute for Physics and Power Engineering, 249033 Obninsk, Russia
\and
           Institut d'Astronomie et d'Astrophysique, ULB, CP226, B1050, Brussels, Belgium 
\and
Institute for Nuclear Power Engineering NRNU MEPHI, 249040 Obninsk, Russia
\and 
Institut fuer Kernphisik, Forschungszentrum Juelich, D-52425 Juelich, Germany}

\abstract{%
   The pygmy-dipole resonances  and photon strength functions in stable and unstable Ni and  Sn isotopes are  calculated 
within the microscopic self-consistent version of the extended theory of finite fermi systems which includes the
QRPA and phonon coupling effects and uses the known Skyrme forces SLy4. The pygmy dipole resonance in $^{72}Ni$ is predicted with the mean energy of 12.4 MeV and the energy-weighted sum rule exhausting 25.6\% of the total strength.
The microscopically obtained  photon E1 strength functions 
are used to calculate nuclear reaction properties, i.e the
radiative neutron capture cross section, gamma-ray spectra, and average radiative widths. Our main conclusion is that in all these quantities  
it is necessary to take the phonon coupling effects into account. }

\maketitle
\section{Introduction}
\label{intro}

During the last decade there has been an increasing interest in the description of the Pygmy Dipole Resonance (PDR) manifested both in ``pure'' low-energy nuclear physics \cite{savran2013,paar2007}
and  in the nuclear data field \cite{gor,ripl2,ripl3} . Usually the PDR is considered in the energy region  between zero and near  the separation neutron energy  and  exhausts typically about 1-2\% of the Energy Weighted Sum Rule (EWSR) but in neutron-rich nuclei , for example $^{68}Ni$ and probably $^{72}Ni$, $^{74}Ni$, this fraction is much larger. It is also  necessary to note that for nuclei with small neutron separation energy, i.e less than 3-4 MeV, the PDR properties significantly change \cite{gor}, and therefore  phenomenological systematics obtained by fitting  characteristics of stable nuclei (with a separation energy of about 8 MeV) is  not suitable. In this sense, phenomenological approaches have no predictive strength and should not be used for applications in nuclear astrophysics.  For these reasons, the microscopic self-consistent approaches, which pretend  (ideally) to describe the ground and excited states of all nuclei  using a small number of  universal parameters like Skyrme interactions or energy density  functionals, have been developed very actively \cite{paar2007,gor}.

As a rule, in the nuclear data field, only phenomenological models for PDR and photon strength functions (PSF), based on various improvements of the Lorentzian-type approximation,  are used (see , in particular, the Reference  Input 
Parameters Libraries, known as RIPL) \cite{ripl2,ripl3}. However,  as it was noted in RIPL2 and RIPL3 \cite{ripl2,ripl3}  , the phenomenological Lorentzian-based   expressions for PSF suffer from various shortcomings: in particular, "they are unable to predict the resonance-like enhancement of the E1 strength at energies below the neutron separation energy" and 
"this approach lacks reliability when dealing with exotic nuclei.'' 
For these reasons,  since 2006,  the microscopic self-consistent PSF calculated within the Hartree-Fock-Bogolyubov method and Quasiparticle Random Phase Approximation  (HFB+QRPA)  have been  included in RIPL2 \cite{ripl2},  RIPL3 \cite{ripl3}  and in  modern nuclear reaction codes  like EMPIRE and TALYS. In general, such an approach  is very natural because it uses the single-particle properties  of each nucleus and is therefore of higher predictive power for the exotic ones in comparison with phenomenological models. However, as we discuss below and as confirmed by modern experiments, the  HFB+QRPA  approach is necessary but not sufficient. To be exact, it should be complemented by the effect describing the interaction of single-particle degrees of freedom with the phonon degrees of freedom, known as the  phonon coupling (PC).

Recent experiments in the PDR energy region have given  new information about the PDR  and PSF structures. 
First,  the  PDR  in the unstable neutron-rich $^{68}$Ni has been found \cite{wiel} at about 11 MeV, which is  higher than the neutron threshold $B_n$ =7.8 MeV, and exhausts about 5\% of the EWSR. Such a measurement was confirmed recently in Ref.~\cite{rossi}. 
Second, experiments on $^{86}$Kr \cite{tsoneva},  Sn isotopes \cite{uts2011,toft} 
and other nuclei clearly  show the 
resonance-like  structures in the PSF and  photo-absorption cross section, 
 which cannot be explained  within the Lorentzian approach.
 Using the known Oslo method, the authors \cite{toft} found out that, in order to explain the observed 
enhancement   of the PSF at E$>$5 MeV with respect to the Generalized Lorentzian description, it is necessary to add a peak structure centered around 8-9 MeV with a strength covering about 2\% of the EWSR.  
 These  structures 
  were explained phenomenologically  by including some additional strength, on top of the HFB+QRPA  predictions, located at a centroid energy of 8.0-8.5 MeV and a strength corresponding to about 1\% of the EWSR \cite{uts2011}.  
  This fact directly confirms  the necessity to improve the HFB+QRPA predictions and search for new sources
  of additional strength. In particular, the PC effects may be at the origin of such an extra strength, as discussed in Refs.\cite{paar2007,revKST}. In our recent article \cite{ave2011},  the self-consistent version of the extended theory of finite fermi systems (ETFFS) \cite{revKST}, in the quasi-particle time blocking approximation (QTBA) \cite{tselyaev} has been used to calculate the standard  strength functions for many  stable and unstable Sn even-even isotopes. This QTBA model includes self-consistently 
the QRPA, phonon coupling and uses a discretized single-particle continuum. 

In the present work,  we use the same theoretical approach to calculate the PSF in  the stable and unstable Ni and Sn isotopes in order :
{\it i)} to describe the PDR in $^{68}$Ni, predict it in $^{72}$Ni  and compare it with the stable $^{58}$Ni isotope,
{\it ii)} to calculate microscopically appropriate PSF and to use them in the reaction code to estimate various nuclear reaction properties,
{\it iii)} to investigate  the  PC contribution to all these quantities.

\section{PSF and PDR in Ni isotopes} 
\label{sec-1}
In all  ETFFS calculations, the standard strength function $S(\omega)=\frac{dB(E(M)L)}{dE} $ is calculated.  
It determines the E1 photoabsorption cross section 
$\sigma(\omega)=4.022\omega S(\omega)$ \cite{revKST},
where the photon energy  $\omega$ is taken in MeV, $S$ is in  fm$^2$MeV$^{-1}$, 
and $\sigma$ is in mb.
 The  PSF is connected to $S(\omega)$ by the simple relation
 \begin{equation}
f(E1) = \frac{1}{3(\pi hc)^2}\frac{\sigma(\omega)}{\omega} = 
  3.487\cdot 10^{-7} S(\omega),
 \end{equation}
 where  $S$ is expressed in  fm$^{2}$MeV$^{-1}$ and  $f(E1)$ in  MeV$^{-3}$.
 
In the present study, we use the SLy4 Skyrme force.  
The ground state is calculated within the  HFB method using the spherical code HFBRAD~\cite{bennaceur}. The residual interaction for the (Q)RPA and  QTBA calculations  is derived as the second derivative of the Skyrme functional. Our smoothing parameter is 200 keV.

\begin{figure}

\centering
\sidecaption
\includegraphics[width=5cm,clip]{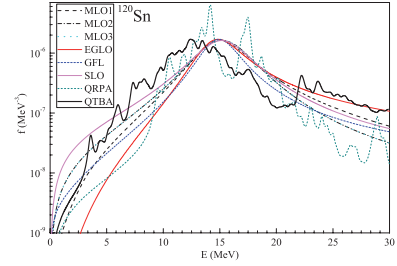}
\caption{ E1 PSF for $^{120}$Sn. Six phenomenological variants from RIPL2 are shown. Dotted line: selfconsistent QRPA. Full line: QTBA (final results with PC) }
\label{fig-1}       
\end{figure}

\begin{figure}

\centering
\sidecaption
\includegraphics[width=5cm,clip]{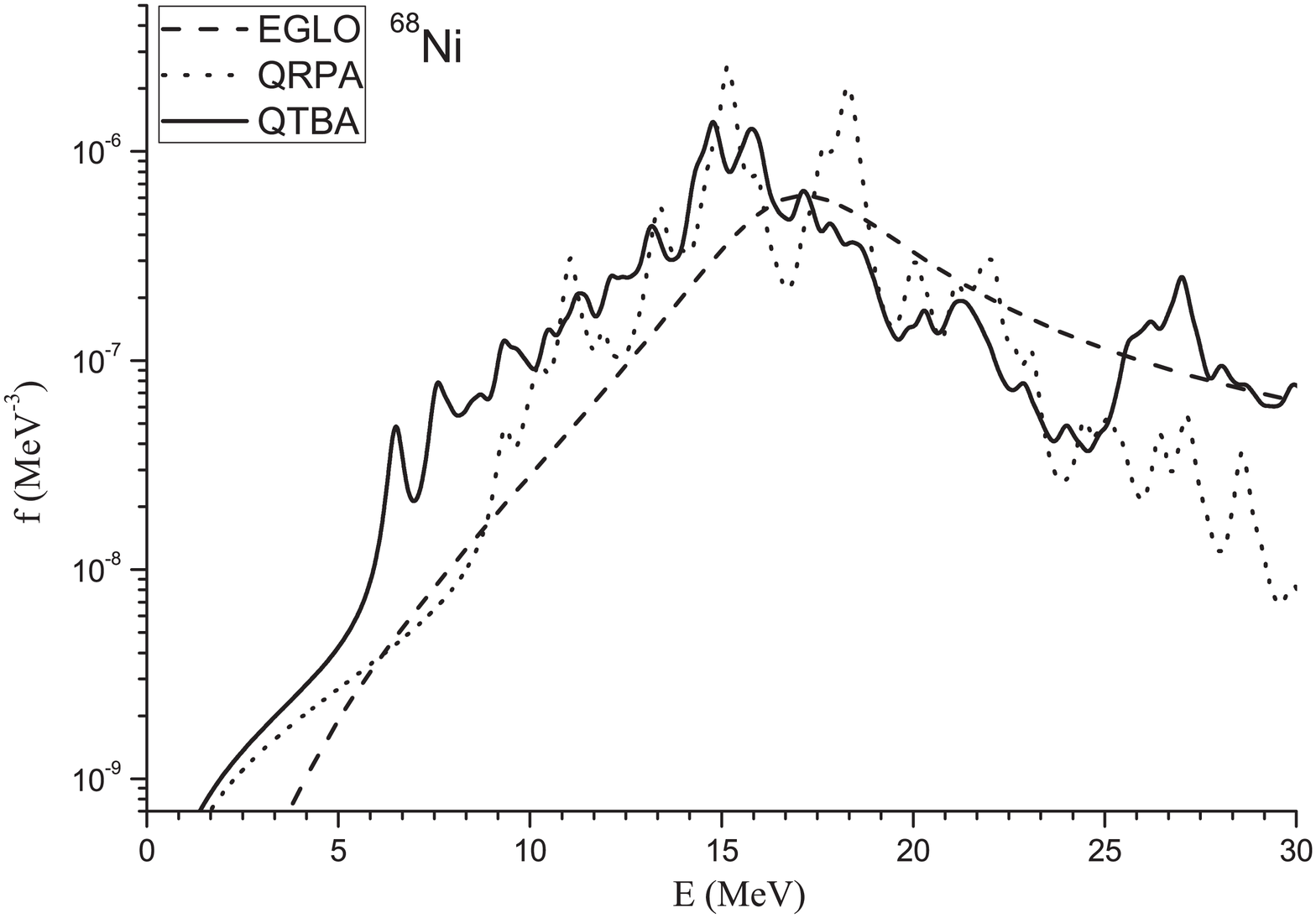}
\caption{E1 PSF for $^{68}$Ni.Dotted line: selfconsistent QRPA. Full line: QTBA (final results with PC). Dashed line : Enhanced Generalized LOrentzian (EGLO) model \cite{ripl2}}
\label{fig-2}       
\end{figure}

\begin{figure}

\centering
\sidecaption
\includegraphics[width=5cm,clip]{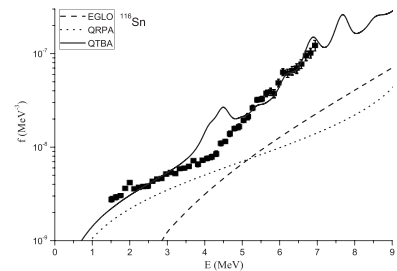}
\caption{Same as in Fig.2 but for $^{116}$Sn. Experiment are taken from
Ref. \cite{toft} }
\label{fig-3}       
\end{figure}

In Fig. \ref{fig-1}, we show  six phenomenological models for the E1 PSF \cite{ripl2} of  $^{120}$Sn, and compare them with our microscopic QRPA and QTBA, i.e. QRPA + PC, results. In Figs. \ref{fig-2} and \ref{fig-3}, the E1 PSF  are shown together with the phenomenological EGLO model \cite{ripl2} for $^{68}$Ni and $^{116}$Sn, respectively. It can be seen that: 1) in  contrast  to phenomenological models, in the Sn and Ni nuclei, structure patterns caused by both the QRPA and PC effects can be observed, the latter  ones  being noticeably lower than the QRPA ones at $E<10$~MeV for Sn isotopes, 2) a good agreement is found with experiment \cite{toft} for $^{116}Sn$ thanks to the inclusion of the PC (Fig. \ref{fig-3}).  

In Table \ref{tab-1}, the integral parameters of the PDR in three Ni isotopes are  given for three PSF models, i.e. the phenomenological EGLO,  our microscopic QRPA and QTBA (QRPA+PC). For comparison,  the 6 MeV interval where the PDR was observed in $^{68}$Ni is considered. In this interval, the PDR characteristics have been approximated, as usual, with a Lorentz curve by fitting the three moments of the Lorentzian and theoretical curves  \cite{revKST}.
A reasonable agreement with experimental data \cite{wiel, rossi} is obtained.
Earlier, a similar calculation was performed for $^{68}$Ni  \cite{lrt2010} using the relativistic QTBA, with  two phonon contributions additionally taking into account.
Concomitantly,  the PDR characteristics in $^{72}Ni$ have been estimated leading in this interval to a mean energy of 12.4 MeV and the large strength of 25.7\% of the total EWSR. In all the isotopes, large PC contribution to the PDR strength is found.

\begin{table}[ht]

\centering
\caption{Integral characteristics of the PDR in Ni isotopes 
calculated in the (8-14) MeV interval for $^{58}$Ni, $^{72}$Ni and (7-13) MeV interval for $^{68}$Ni (see text for details).}
\label{tab-1}
\begin {tabular}{ l l l l l l l l}
\hline
\multirow{2}{*}{Nuclei}&\multicolumn{2}{l}{EGLO}&\multicolumn{2}{l}{QRPA}&\multicolumn{2}{l}{QTBA}\\
\cline{2-7}
&E,MeV&\%&E,MeV&\%&E,MeV&\%\\
\hline
$^{58}$Ni&11.3&2.4&13.3&6&14.0&11.7\\
\hline
$^{68}$Ni&11.3&3.1&11.0&4.9&10.8&8.7\\
\hline
$^{72}$Ni&11.3&3.4&12.4&14.7&12.4&25.7\\
\hline
\end{tabular}
 \end{table}

\section{Neutron radiative capture }
\label{sec-2}
\begin{figure}

\centering
\sidecaption
\includegraphics[width=5cm,clip]{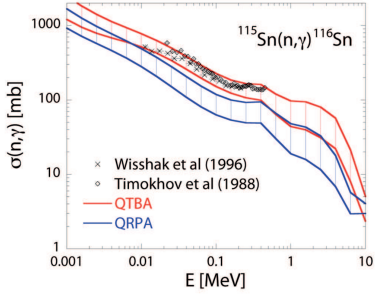}
\caption{$^{115}$Sn(n,$\gamma$) cross section calculated with the  QRPA (blue) and QTBA (red) PSF. The uncertainty bands depict the uncertainties affecting the nuclear level density predictions \cite{kon08,gor08,hil12}.
Experimental cross section are taken from Refs.~\cite{wisshak,timokhov}}
\label{fig-4}       
\end{figure}

In Fig. \ref{fig-4}, the   radiative neutron capture cross section  for the $^{115}$Sn obtained  with the QRPA and QRPA+PS photon E1 strength functions is shown. We also show  the uncertainty bands obtained when considering various nuclear level  density models \cite{kon08,gor08,hil12}. One can see clearly that the agreement with experiment is possible only when the  PC is taken into account. 

\begin{figure}

\centering
\sidecaption
\includegraphics[width=5cm,clip]{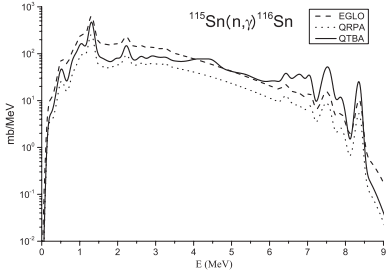}
\caption{Gamma-ray spectra from $^{115}Sn(n,\gamma)$ for the neutron energy of 100 keV
}
\label{fig-5}      
\end{figure}

The corresponding  capture gamma-ray spectra calculated for the neutron  energy of 100 keV are given in Fig. \ref{fig-5} and compared with the result obtained with the EGLO PSF model. Here the phenomenological Generalized Superfluid model \cite{ripl2} of nuclear level densities is adopted. On the whole, our results are in an agreement with the EGLO ones (for this stable nucleus) and show that the  PC contribution is significant. For all three variants some  structures are found. Unfortunately, no experimental data exist for this case.

\section{Average  radiative widths }
\label{sec-3}

Average radiative widths of neutron resonances $\Gamma_{\gamma}$ are very important properties of  gamma-decay from high-energy nuclear states; they  are used in nuclear reaction calculations, in particular to normalise the $\gamma$-ray strength.  There are a lot of  experimental data for them \cite{ripl1,muhab}. For 11 Sn and Ni isotopes, we have calculated the widely used quantity 
$2\pi\Gamma_{\gamma}/D_0$ with the EGLO and our QRPA and QTBA PSF models and the GSM nuclear level density model \cite{ripl2} for the s-wave spacing $D_0$ 
(Table \ref{tab-2}).  As far as we know, these are the first calculations performed with PC. We found that, except for $^{68}$Ni,   the  PC increases  the QRPA contribution by  100-200\%  in  the direction of the  systematics. The results for $^{120}$Sn and $^{62}$Ni, for which experimental data (not systematics) exists,  are of great interest. Here we obtain a good agreement with experiment, which probably means that the M1 contribution is not large (especially for $^{120}$Sn). As far as the EGLO model is concerned, it is clear that the predictions  differ from the systematics \cite{ripl1} and experimental data (and also from our QTBA results)  rather  strongly. 

\begin{table}[ht]

\centering
\caption{Quantities $2\pi\Gamma_{\gamma}/D_0$ for s-neutrons where $\Gamma_{\gamma}$ is the average radiative width (see  text for details). The experimental data (underlined) and systematics are taken from \cite{muhab} and \cite{ripl1}, respectively.}
\label{tab-2}
\begin {tabular}{ l l l l l }
\hline
Nuclei&EGLO&QRPA&QTBA&\underline{exp.} or system.\\
\hline
$^{110}$Sn&$12.6\cdot10^{-2}$&	$3.90\cdot10^{-2}$&	$7.99\cdot10^{-2}$&	$9.57\cdot10^{-2}$\\
\hline
$^{112}$Sn&$9.44\cdot10^{-2}$&	$3.08\cdot10^{-2}$&	$5.88\cdot10^{-2}$&	$9.76\cdot10^{-2}$\\
\hline
$^{116}$Sn&$7.99\cdot10^{-3}$&	$3.33\cdot10^{-3}$&	$5.79\cdot10^{-3}$&	$11.73\cdot10^{-3}$\\
\hline
$^{120}$Sn&$5.77\cdot10^{-3}$&	$2.52\cdot10^{-3}$&	$7.10\cdot10^{-3}$&	\underline{$6.98\cdot10^{-3}$}\\
\hline
$^{124}$Sn&$4.77\cdot10^{-3}$&	$2.13\cdot10^{-3}$&	$2.67\cdot10^{-3}$&	$9.84\cdot10^{-3}$\\
\hline
$^{132}$Sn&$6.53\cdot10^{-4}$&	$2.18\cdot10^{-4}$&	$2.42\cdot10^{-4}$&	$1.39\cdot10^{-4}$\\
\hline
$^{136}$Sn&$9.90\cdot10^{-7}$&	$9.97\cdot10^{-7}$&	$1.10\cdot10^{-6}$&	$6.51\cdot10^{-6}$\\
\hline
$^{58}$Ni&$7.04\cdot10^{-3}$&	$2.30\cdot10^{-3}$&	$7.33\cdot10^{-3}$&	$17.0\cdot10^{-3}$\\
\hline
$^{62}$Ni&$2.51\cdot10^{-3}$&	$1.97\cdot10^{-3}$&	$4.33\cdot10^{-3}$&\underline{$5.98\cdot10^{-3}$}\\
\hline
$^{68}$Ni&$1.04\cdot10^{-4}$&	$4.73\cdot10^{-5}$&	$2.46\cdot10^{-4}$&	$2.64\cdot10^{-4}$\\
\hline
$^{72}$Ni&$5.08\cdot10^{-5}$&$1.33\cdot10^{-5}$&$2.45\cdot10^{-5}$&$5.08\cdot10^{-5}$\\
\hline	
\end{tabular}
 \end{table}

\section{Conclusion}
\label{sec-4}

The  characteristics of nuclear reactions with gamma-rays have been calculated within the  microscopic self-consistent approach which takes into account the QRPA and PC effects and uses the SLy4 Skyrme force. Such a self-consistent approach is of particular relevance for nuclear astrophysics. A reasonable  agreement with experiment  has been obtained for the PSF in $^{116}$Sn and  the PDR integral properties  in $^{68}Ni$. We predict the PDR in the spherical $^{72}Ni$ nucleus  at 12.4 MeV with a very large strength corresponding to 25.7\% of the EWSR. For the first time, the average radiative widths 
have been calculated microscopically with the  PC taken into account. In all the considered quantities, the contribution of PC turned out to be significant. These results confirm the necessity to  include the PC effects  into the theory of nuclear data both for stable and unstable nuclei.
 
\vspace{0.5cm}
S. Kamerdzhiev acknowledges the CGS15 Organizing Committee and Prof. R. Schwengner for the invitation and support,  Drs. V. Furman and A. Sukhovoj for discussions of the results.

\end{document}